\documentclass[aps,pra,showpacs,superscriptaddress,twocolumn,
amsmath,amssymb,longbibliography]{revtex4-1}
\usepackage{amsfonts}
\usepackage{txfonts}
\usepackage{amsmath}
\usepackage{float}
\usepackage{diagbox}
\usepackage[colorlinks,breaklinks,linkcolor=blue,anchorcolor=blue,citecolor=blue,urlcolor=blue]{hyperref}
\usepackage{graphicx}
\usepackage{dcolumn}
\usepackage{bm}
\usepackage{amssymb}
\usepackage{mathrsfs}
\usepackage[sort&compress]{natbib}
\usepackage{subfigure}
\usepackage{braket}

\begin{document}
\title{Generation of high-fidelity Greenberger-Horne-Zeilinger states in a driven hybrid quantum system}
\author{Xin Zeng}
\affiliation{Center for Quantum Sciences and School of Physics, Northeast Normal University, Changchun 130024, China}

\author{Yuxin Kang}
\affiliation{Center for Quantum Sciences and School of Physics, Northeast Normal University, Changchun 130024, China}

\author{Chunfang Sun}
\affiliation{Center for Quantum Sciences and School of Physics, Northeast Normal University, Changchun 130024, China}

\author{Chunfeng Wu}
\affiliation{Science, Mathematics and Technology, Singapore University of Technology and Design, 8 Somapah Road, Singapore 487372, Singapore}

\author{Gangcheng Wang}
\email{wanggc887@nenu.edu.cn}
\affiliation{Center for Quantum Sciences and School of Physics, Northeast Normal University, Changchun 130024, China}
\date{\today}

\begin{abstract}
In this study, we propose a theoretical scheme for achieving long-distance Greenberger-Horne-Zeilinger states in a driven hybrid quantum system. By applying a microwave field to the YIG sphere, we utilize the Kerr effect to induce the squeezing of the magnon, thereby achieving an exponential enhancement of the coupling strength between the magnonic mode and spins, and we also discuss in detail the relationship between the squeezing parameter and the external microwave field. By means of the Schrieffer-Wolff transformation, the magnonic mode can be adiabatically eliminated under the large detuning condition, thereby establishing a robust effective interaction between spins essential for realizing the desired entangled state. Numerical simulations indicate that the squeezing parameter can be effectively increased by adjusting the driving field, and our proposal  can generate high-fidelity Greenberger-Horne-Zeilinger states even in dissipative systems. Additionally, we extensively discuss the influence of inhomogeneous broadening on the entangled states, and the experimental feasibility shows that our results provide possibilities in the realms of quantum networking and quantum computing.
\end{abstract}
\maketitle

\allowdisplaybreaks
\section{Introduction}
\label{Sec:I}
Quantum entanglement \cite{Schrodinger1935,PhysRevA.67.022112,PhysRevA.65.032108,PhysRevA.95.062105,Katz2022}, as a valuable resource for quantum information \cite{6Lambropoulos2007,7Vedral2010,8Mirrahimi2014}, has garnered significant attention due to its crucial role in quantum computing \cite{9Imai2006,10Nielsen2010}, quantum metrology \cite{11PhysRevA.66.023819,12RevModPhys.90.035005,13Li2023} and other fields. Recently, a wide range of potential applications has been demonstrated in across various systems, including superconducting circuit \cite{14PRXQuantum.2.017001,15PhysRevLett.119.180511,16PhysRevE.104.054211,17PhysRevLett.129.037205}, photons \cite{18PhysRevA.62.032301,19PhysRevLett.115.023602,20PhysRevA.107.052614}, trapped ions \cite{21PhysRevLett.106.130506,22Li2023,23e24060813}, atomic ensembles \cite{24Li2023,25PhysRevLett.130.030602,26PhysRevLett.127.093602} and solid material \cite{NVLi2023,28Qin_2024}. Overall, these approaches involving high-dimensional entangled states have introduced more precise operations and complex correlations, which can enable the generation of remarkable coherent characteristics and a robust exchangeable scheme. However, the strategies  that rely on global control of the spins typically require sensitive measurement-based feedback, high-fidelity manipulation, and accurate preparation times, resulting in costly quantum gate operations and extensive coherent control. Consequently, the major problem that needs to be addressed is achieving long-distance and highly robust quantum information exchange.

Greenberger-Horne-Zeilinger (GHZ) states, originally regarded as the paradigmatic example of maximal entanglement \cite{29PhysRevLett.65.1838}, represents a special type of multipartite entangled state that violates the Bell inequality to the greatest extent. Due to its favorable properties for quantum information theory, GHZ states shows promise in achieving reliable quantum information processing. Generally speaking, compared with other states typically used in quantum computation, GHZ states exhibit a range of potential advantages, including strong multipartite correlations, non-classical properties, and robustness against various sources of noise. Up to now, several studies have demonstrated collaborative and dedicated efforts towards achieving the generation of high-fidelity GHZ states \cite{32PhysRevA.53.4591,33PhysRevLett.87.230404,34PhysRevA.82.062326,35Izadyari2016,36Li2024,37PhysRevApplied.18.064036}, which have been successfully implemented on various platforms such as hybrid cavity-magnon systems, optical tweezer arrays, circuit QED, and ion clusters. One approach that has attracted considerable attention involves employing trapped ions as qubits, which allows for the use of laser cooling techniques to ensure stability and coherence between them. Nevertheless, these previous studies have demonstrated high-fidelity GHZ states on a smaller scale but have failed to address external interference noise and increased preparation time. The scalability of these methods and long-distance entangled states remains an open question, which is crucial for advancing applications in efficient quantum algorithms and multi-party quantum communication \cite{31PhysRevLett.129.070601}.

In parallel, magnon stands as collective microwave excitations of electron spins in ferromagnetic materials \cite{49PhysRevLett.127.087203}, bringing significant potential for exploring macroscopic
quantum effects \cite{49PhysRevLett.127.087203}. Its remarkable nonlinear property enables the establishment of an efficient quantum medium for inducing spin-spin interactions under the condition of coherent manipulation \cite{50PhysRevLett.125.247702}. In particular, indirect interactions can overcome the limitations of direct coupling for alleviating spatial constraints, thereby further facilitating the feasibility of GHZ states on a larger scale. On the other hand, nitrogen vacancy (NV) centers in diamond have been widely recognized as exceptional solid-state qubits due to their prolonged coherence times and remarkable controllability. Another distinct feature of these systems is that the interaction between the NV spins arising from the relative motion and a source of local magnetic-field gradients can arise extrinsically or intrinsically. Unfortunately, the results of induced interaction have received little attention regarding their coupling strength, which poses a significant obstacle to achieving strong coupling at the quantum level. A natural question arises as to whether the strong coupling mediated by magnons and NV centers can be harnessed to further achieve the preparation of entangled states. This motivates us to delve into the dynamics of such hybrid quantum systems.

In this paper, we present a theoretical proposal for the preparation of high-fidelity GHZ states over an extended distance scale. By utilizing the Kerr nonlinear term \cite{55PhysRevB.108.024105,56PhysRevB.107.144411} within a strong external field, our method involves the direct weak coupling between a magnon and multiple NV centers, which can significantly enhance the coupling strength, thereby ensuring effective squeezing of the magnonic state. Furthermore, we adiabatically eliminate the magnonic mode \cite{57PhysRevA.102.042605} to derive the effective spin-spin interaction, which allows for the preparation of GHZ states at a specific time \cite{58Li2019}. The description of numerical simulation results reveal that our scheme can achieve extremely reliable GHZ states even in the presence of dissipation. Additionally, we introduce the so-called cavity protection mechanism \cite{59PhysRevA.83.053852} to mitigate the effect of inhomogeneous broadening and give a thorough explanation for experimental feasibility.

This paper is structured as follows. In Sec. \ref{Sec:II}, a sequence of transformations has been applied to derive the effective Hamiltonian with exponential-scale coupling strength. In Sec. \ref{Sec:III}, we show the detailed scheme for preparing the GHZ states in the dispersion regime, and analyze the impact of dissipation on the dynamic evolution process. In Sec. \ref{Sec:IV}, we also consider the effect of cavity protection to eliminate the influence of inhomogeneous broadening. Finally, we provide the experimental feasibility for our scheme and give a comprehensive summary of the entire paper in Sec. \ref{Sec:V}.
\begin{figure}[t]
\centering
\subfigure{
\includegraphics[width=0.75\columnwidth]{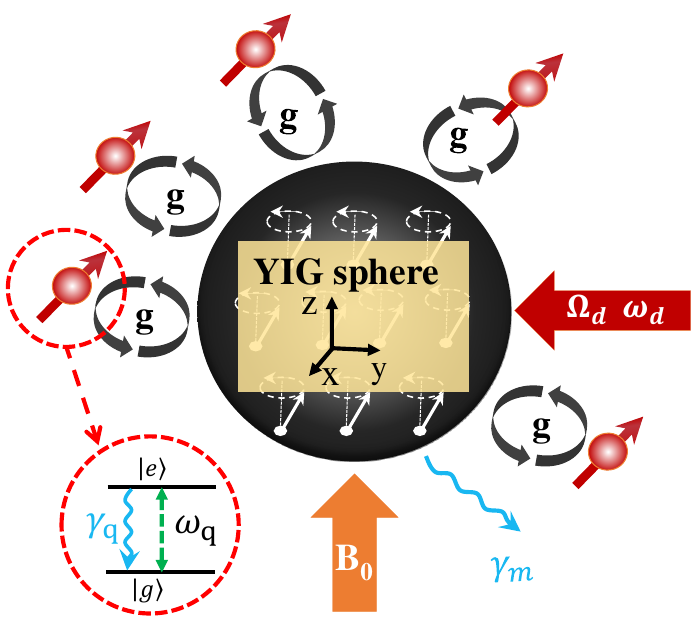}}
\caption{Diagram of a hybrid system with multiple NV center spins weakly coupled to a Kerr magneton of radius $R$. Each NV center, characterized by excited state $\ket{e}$ and ground state $\ket{g}$ is located $d$ distance from the sphere surface, and the magnon is driven by a strong driving field with amplitude $\Omega_{d}$ and frequency $\omega_{d}$.}
\label{Fig.1}
\end{figure}
\section{Model and Hamiltonian}
\label{Sec:II}
We consider a hybrid quantum system consisting of multiple NV centers as spin qubits with the transition frequency $\omega_{q}$, weakly coupled to the Kerr magnon within a nanometer-scale radius $R$ of the yttrium-iron-garnet (YIG) sphere. Each NV center spin is positioned at a distance $d$ from the surface of the sphere, while the magnon is driven by a microwave field with frequency $\omega_{d}$ and amplitude $\Omega_{d}$ as illustrated in Fig. \ref{Fig.1}. In this case, taking into account the Kerr nonlinearity \cite{50PhysRevLett.125.247702} of the magnon, the total Hamiltonian of the hybrid system under the rotating-wave approximation (RWA) is (setting $\hbar=1$)
\begin{equation}\label{Eq:1}
    \hat{H}_{T}(t)=\hat{H}_{q}+\hat{H}_{K}+\hat{H}_{V}+\hat{H}_{d}(t),
\end{equation}
 where
 \begin{subequations}\label{Eq:2}
  \begin{align}
      &\hat{H}_{q}=\omega_{q}\hat{J}_{z},\\
    &\hat{H}_{K}=\omega_{m}\hat{m}^{\dagger}\hat{m}-\frac{K}{2}\hat{m}^{\dagger}\hat{m}^{\dagger}\hat{m}\hat{m},\\  &\hat{H}_{\rm int}=g(\hat{J}_{+}\hat{m}+\hat{J}_{-}\hat{m}^{\dagger}),\\
       &\hat{H}_{d}(t)=\Omega_{d}(\hat{m}^{\dagger}e^{-i\omega_{d}t}+\hat{m}e^{i\omega_{d}t}).
  \end{align}   
 \end{subequations}
Here the collective operation $\hat{J}_{\alpha}=\sum^{N}_{j=1}\hat{\sigma}^{\alpha}_{j}/2$ and $\hat{J}_{\pm}=\hat{J}_{x}\pm i\hat{J}_{y}$ with $\hat{\sigma}^{\alpha}_{j}\ (\alpha=x,y,z)$ being the spin operators acting on the $j$-th site around the YIG sphere. The collective operators satisfy the $\mathfrak{su}(2)$ algebra relation: $[\hat{J}_{\alpha}, \hat{J}_{\beta}]= i\epsilon_{\alpha\beta\gamma}\hat{J}_{\gamma}$ with $\epsilon_{\alpha\beta\gamma}$ being Levi-Civita symbol. $\hat{m}^{\dagger} (\hat{m})$ is the creation (annihilation) operator of the kittle mode with the frequency $\omega_{m}$. The Kerr coefficient $K=2 \mu_{0} K_{\rm{an}} \gamma^{2} /\left(M^{2} V_{m}^{2}\right)$ characterizes the anharmonicity of the magnon, where $\mu_{0}$ is the vacuum permeability, $K_{\rm{an}}$ is the first-order anisotropy constant of the magnon, $M$ is saturation magnetization, $V_{m}$ is the volume of the YIG sphere and the gyromagnetic ratio $\gamma /2\pi=g_{e}\mu _{e}/\hbar$ is defined, with $g_{e}$ being the $g$ factor and $\mu_{e}$ representing the Bohr magneton \cite{42PhysRevB.105.245310}.

For a deeper comprehension of such a system, we can derive the following Hamiltonian based on the rotating frame with respect to the driving frequency $\omega_{d}$
\begin{equation}\label{Eq:3}
\hat{H}_{T}'=\Delta_{q}\hat{J}_{z}+\delta_{m}\hat{m}^{\dagger}\hat{m}-\frac{K}{2}\hat{m}^{\dagger2}\hat{m}^{2}+g(\hat{J}_{+}\hat{m}+\hat{J}_{-}\hat{m}^{\dagger})+\Omega_{d}(\hat{m}^{\dagger}+\hat{m}),
\end{equation}
where $\Delta_{q}=\omega_{q}-\omega_{d}$, $\delta_{m}=\omega_{m}-\omega_{d}$. Taking into account the dissipative effect of the system in the driving frequency framework, the Heisenberg-Langevin equation describing the dynamics of the coupled magnonic mode can be expressed as follows \cite{56PhysRevB.107.144411}
\begin{equation}\label{Eq:4}
    \begin{aligned}
 \dot{\hat{m}}=&-i(\delta_{m}-i\frac{\gamma_{m}}{2})\hat{m}+iK\hat{m}^{\dagger}\hat{m}^{2}-ig\hat{J}_{-}-i\Omega_{d}+\hat{M}_{\rm{in}},
    \end{aligned}
\end{equation}
where $\gamma_{m}$ and $\hat{M}_{\rm{in}}$ denote the decay rate and noise operator of the magnon. Then, we derive the linearized dynamics of quantum fluctuations around the steady-state expectation values of the system. The operator in Eq. (\ref{Eq:4}) can be decomposed into the sum of the steady-state value and quantum fluctuations, i.e., $\hat{A}=\braket{\hat{A}}+\delta\hat{A}$. Substituting this expansion into Eq. (\ref{Eq:4}) allows us to obtain the steady-state value as follows
\begin{equation}\label{Eq:5}
\begin{split}
           \braket{\hat{m}}\approx\frac{\Omega_{d}}{K|\braket{\hat{m}}|^{2}-\delta_{m}+i\frac{\gamma_{m}}{2}}.
\end{split}
\end{equation}
\begin{figure*}[htbp]
\centering
\subfigure{
\includegraphics[width=0.9\textwidth]{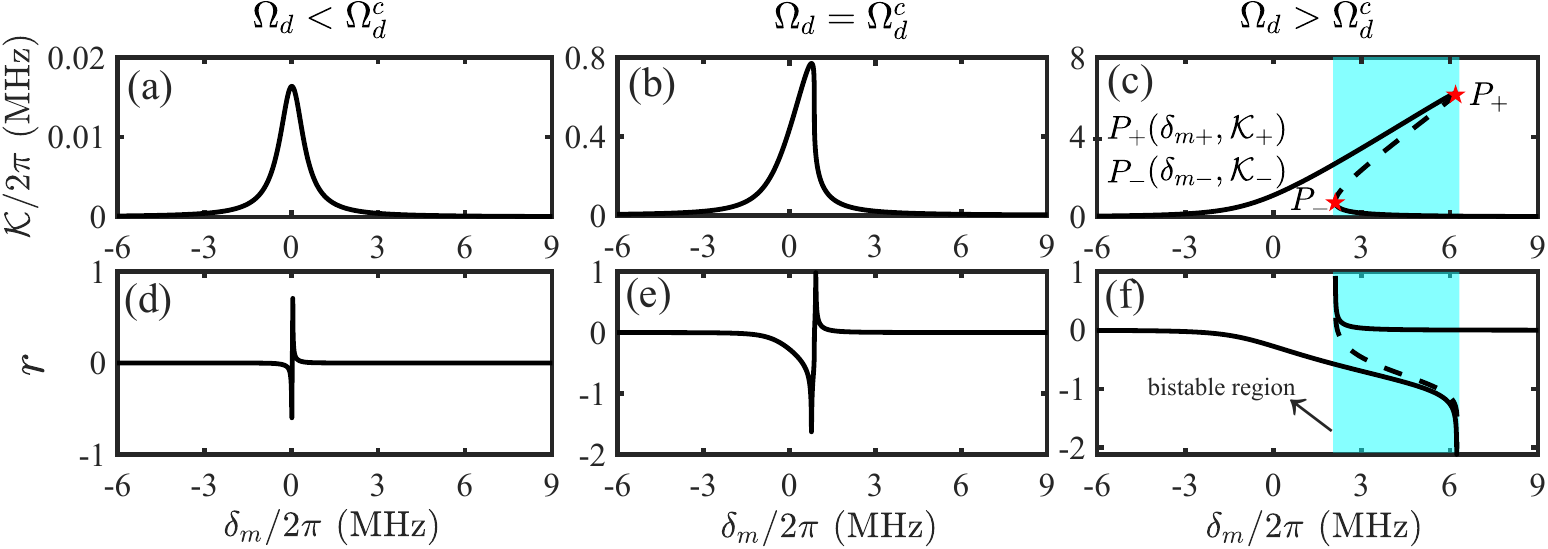}}
\caption{Panels (a)-(c) illustrate the relationship between $\mathcal{K}$ and the detuning $\delta_{m}$, while Panels (d)-(f) show the relationship between $r$ and the detuning $\delta_{m}$ under different driving amplitudes. Figures (a) and (d) correspond to $\Omega_{d}<\Omega_{d}^{c}$, Figures (b) and (e) represent  $\Omega_{d}=\Omega_{d}^{c}$, and Figures (c) and (f) correspond to $\Omega_{d}>\Omega_{d}^{c}$. The stable and unstable parts are indicated by solid and dashed lines, and the blue region represents the bistability area while $P_{\pm}$ denotes the coordinates of the bistability critical points. We take $g=2\pi\times0.5$ MHz, $\gamma_{m}=2\pi\times1$ MHz, $\gamma_{q}=2\pi\times1$ kHz, $K=2\pi\times1.56\ \mu$Hz, and $\Delta_{q}=2\pi\times6$ GHz.}
\label{Fig.2}
\end{figure*}
Otherwise, it is worth mentioning that in the presence of strong driving conditions, the steady-state average value of the system is particularly large, which results in an extremely weak fluctuation amplitude associated with the steady-state value of the operator, which allows to safely disregard higher-order fluctuation components \cite{61PhysRevLett.98.030405}. Consequently, by inserting the above expression into Eq. (\ref{Eq:4}), we can obtain the quantum Langevin equations for the fluctuation operator 
\begin{equation}\label{Eq:6}
\begin{split}
      \delta\dot{\hat{m}}=&-i(\delta_{m}-i\frac{\gamma_{m}}{2})\delta\hat{m}+2iK|\braket{\hat{m}}|^{2}\delta\hat{m}+iK\braket{\hat{m}}^{2}\delta\hat{m}^{\dagger}\\
      &-ig\delta\hat{J}_{-}+\hat{M}_{\rm{in}}.
\end{split}
\end{equation}
By rewriting the equation of motion in Eq. (\ref{Eq:6}) as $\delta\dot{\hat{m}}=-i[\delta\hat{m},\hat{H}_{L}]-\frac{\gamma_{m}}{2}\delta\hat{m}+\hat{M}_{\rm{in}}$, we obtain the Hamiltonian after eliminating non-linear Kerr term 
\begin{equation}\label{Eq:7}	\hat{H}_{L}=\Delta_{q}\hat{J}_{z}+\Delta_{m}\hat{m}^{\dagger}\hat{m}-\frac{1}{2}(\mathcal{K}\hat{m}^{2}+\mathcal{K}^{*}\hat{m}^{\dagger2})+g(\hat{J}_{+}\hat{m}+\hat{J}_{-}\hat{m}^{\dagger}),
\end{equation}
with $\Delta_{m}=\omega_{m}-\omega_{d}-2K|\braket{\hat{m}}|^{2}$ representing the effective magnon-number-dependent frequency detuning, and $\mathcal{K}=K\braket{\hat{m}}^{2}$ denoteing the enhanced coefficient of the two-magnon process. Meanwhile, we have the capability to precisely manipulate the crystallographic axis of the YIG sphere, thereby ensuring that $K$ can experimentally take on either a positive or a negative value. On the other hand, a potentially larger Kerr coefficient may exist for the YIG nanosphere, along with the adjustability of $\braket{\hat{m}}$ through the applied driving field, which can give rise to the magnon squeezing required in the approach. Due to the favorable property in the above Hamiltonian, we consider the two-magnon process to employ the squeezed-magnon framework via the transform $\hat{U}=\exp[r(\hat{m}^{2}-\hat{m}^{\dagger2})/2]$ with the squeezing parameter satisfies: $r=\frac{1}{4}\ln \left[\left(\Delta_{m}+\mathcal{K}\right) /\left(\Delta_{m}-\mathcal{K}\right)\right]$. Substituting these expression into Eq. (\ref{Eq:7}), the Hamiltonian under the squeezed-magnon framework can be recast as
\begin{equation}\label{Eq:8}
    \begin{aligned}	
    \hat{H}_{S}=&\hat{U}^{\dagger}\hat{H}_{L}\hat{U}\\
    =&\Delta_{q}\hat{J}_{z}+\frac{\Delta_{m}}{\cosh2r}\hat{m}^{\dagger}\hat{m}+ge^{r}(\hat{J}_{+}+\hat{J}_{-})(\hat{m}+\hat{m}^{\dagger})\\
 &+ge^{-r}(\hat{J}_{+}-\hat{J}_{-})(\hat{m}-\hat{m}^{\dagger}),   
    \end{aligned}	
\end{equation}
which features two effective coupling strengths, both exhibiting exponential positive or negative growth as $r$ increases. It is evident that the variation of $r$ is related to the values of the driving field frequency $\omega_{d}$ and $\mathcal{K}$. Under the steady-state conditions, $\mathcal{K}$ satisfies the following cubic nonlinear equation (more detailed derivations are reported in Appendix \ref{APPENDIX:A})
\begin{equation} \label{Eq:9} \mathcal{K}^{3}+C_{2}\mathcal{K}^{2}+C_{1}\mathcal{K}+C_{0}=0,
\end{equation}
where $C_{0}=-K\Omega_{d}^{2}$, $C_{1}=\delta_{m}^{2}+\gamma_{m}^{2}/4$, and $C_{2}=-2\delta_{m}$. The cubic nonlinear Eq. (\ref{Eq:9}) can be understood as a function of $\mathcal{K}$ with respect to $\delta_{m}$, revealing the bistability of $\mathcal{K}$ induced by the Kerr effect while driving the magnetic resonator. The region where bistability occurs corresponds to the case where $\mathcal{K}$ has three unequal real roots, and we will provide a detailed discussion on the extent of the region where bistability occurs. For the critical boundary of bistability, the following condition is satisfied
\begin{equation}\label{Eq:10}
    \frac{{\rm d}\delta_{m}}{{\rm d}\mathcal{K}}=0.
\end{equation}
Additionally, Eq. (\ref{Eq:9}) can be regarded as a bivariate implicit function $F (\mathcal{K},\delta_{m})=0$ with $F (\mathcal{K},\delta_{m})=\mathcal{K}^{3}+C_{2}\mathcal{K}^{2}+C_{1}\mathcal{K}+C_{0}$. According to the implicit function theorem
\begin{equation}\label{Eq:11}
    \frac{{\rm d}\delta_{m}}{{\rm d}\mathcal{K}}=-\frac{\partial_{\mathcal{K}}F(\mathcal{K},\delta_{m})}{\partial_{\delta_{m}}F(\mathcal{K},\delta_{m})}.
\end{equation}
By combining Eq. (\ref{Eq:10}), we can derive the critical condition
\begin{equation}\label{Eq:12}
\begin{aligned}
    &\partial_{\mathcal{K}} F=3\mathcal{K}^{2}+2C_{2}\mathcal{K}+C_{1}=0.
\end{aligned}
\end{equation}
When the quadratic equation in Eq. (\ref{Eq:12}) has two equal real roots, there is a bistable critical point for this system. According to the discriminate of the Eq. (\ref{Eq:12}), we obtain the critical point satisfy the following critical conditions
\begin{equation}\label{Eq:13}
        \delta_{m}^{c}=\frac{\sqrt{3}}{2}\gamma_{m},\quad \mathcal{K}^{c}=\frac{\sqrt{3}}{3}\gamma_{m}.
\end{equation}
Substituting Eq. (\ref{Eq:13}) into Eq. (\ref{Eq:9}), we can obtain the critical driving field amplitude $\Omega_{d}^{c}=\sqrt{\sqrt{3}\gamma_{m}^{3}/9K}$. Obviously, when the driving field amplitude satisfies: $\Omega_{d}>\Omega_{d}^{c}$, the system exhibits bistability. The two real roots of Eq. (\ref{Eq:12}) correspond to two saitching points of bistable regions
\begin{equation}\label{Eq:14}
\begin{aligned}
    \mathcal{K}_{+}=&\frac{1}{3}\left( 2\delta_{m+}+\sqrt{\delta_{m+}^{2}-\frac{3}{4}\gamma_{m}^{2}}\right) ,\\
    \mathcal{K}_{-}=&\frac{1}{3}\left( 2\delta_{m–}-\sqrt{\delta_{m–}^{2}-\frac{3}{4}\gamma_{m}^{2}}\right).
\end{aligned}
\end{equation}
Thus, for a specified amplitude $\Omega_{d}$ of the driving field, the saitching points of the bistable state $(\delta_{m-},\delta_{m+})$ can be precisely identified, along with the associated range of $(\mathcal{K}_{-}, \mathcal{K}_{+})$. Under the condition of the steady state, the Hamiltonian not only allows for effective linearization, but the squeezing parameter $r$, which is dependent on $\mathcal{K}$ and $\delta_{m}$, can also be optimized by varying the frequency of the driving field. To provide an intuitive illustration of the bistable behavior of $\mathcal{K}$ induced by the Kerr effect in a steady state, as well as its monostable behavior under different $\Omega_{d}$, we present $\mathcal{K}$ and $r$ as functions of the magnetic oscillator detuning $\delta_{m}$ in Fig. \ref{Fig.2}, while taking into account $K=2\pi\times1.56\ \mu$Hz for a 40-$\mu$m-diameter YIG sphere \cite{PhysRevResearch.3.023126}, and the other experimental parameters \cite{Wang2023,PhysRevA.109.022442,PhysRevA.108.063703,42PhysRevB.105.245310}: $g=2\pi\times0.5$ MHz, $\gamma_{m}=2\pi\times1$ MHz, $\gamma_{q}=2\pi\times1$ kHz, and $\Delta_{q}=2\pi\times6$ GHz. Obviously, when $\Omega_{d}\leqslant\Omega_{d}^{c}$, the system is in a monostable state, whereas when $\Omega_{d}>\Omega_{d}^{c}$, it transitions to a bistable state. The blue region corresponds to the area where bistability occurs, and $P_{\pm}(\delta_{m\pm},\mathcal{K}_{\pm})$ (red pentagons) denote the two saitching points of bistability. Notably, by employing the Routh-Hurwitz criteria \cite{PhysRevResearch.5.043053}, the stability of the system can be effectively assessed, and under the chosen parameter conditions, there are two stable branches (solid lines) and one unstable middle branch (dashed lines). Furthermore, bistability only occurs when the detuning $\delta_{m}$ is significantly smaller than its own natural frequency $\omega_{m}$ \cite{PhysRevA.108.023503}. Additionally, it is worth noting that $P_{-}(\delta_{m-},\mathcal{K}_{-})$ and $P_{+}(\delta_{m+},\mathcal{K}_{+})$ correspond to the maximum and minimum values of $r$, respectively. This directly leads to an enhanced spin-magnon coupling that is larger than the original strength. With such method, the separation between YIG sphere and spins can be increased when maintaining the same large spin-magnon coupling \cite{42PhysRevB.105.245310,QIN20241}. 
\section{Preparation of Multi-Qubit GHZ States}
\label{Sec:III}
\subsection{Magnon-induced spin-spin interaction}
In this section, we will provide a detailed explanation of how to generate the one-axis twisting (OAT) interaction, which paves the way for the subsequent preparation of multi-particle GHZ entangled states. Building upon the preceding contents, we can increase the squeezing parameter $r$ to ensure the condition $e^{r}\gg 1$, making $e^{-r}$ approach zero, which allows us to safely ignore the term $\hat{H}_{Sq}=ge^{-r}(\hat{J}_{+}-\hat{J}_{-})(\hat{m}-\hat{m}^{\dagger})$. In this case, the Hamiltonian (\ref{Eq:8}) reads
\begin{equation}\label{Eq:15} \hat{H}'_{S}=\Delta_{q}\hat{J}_{z}+\widetilde{\omega}_{m}\hat{m}^{\dagger}\hat{m}+G(\hat{J}_{+}+\hat{J}_{-})(\hat{m}+\hat{m}^{\dagger}),
\end{equation}
where $\widetilde{\omega}_{m}=\Delta_{m}/\cosh2r$ with $G=ge^{r}$. For the dispersion regime, i.e., $G/\Delta_{\pm}\ll 1$ with $\Delta_{\pm}=\Delta_{q}\pm\widetilde{\omega}_{m}$, we can further employ the so-called Schrieffer-Wolff (SW) transformation \cite{64PhysRevA.102.042605} to adiabatically eliminate the magnonic mode, and obtain the effective strong interaction between long distant spins. In this method, a small expansion parameter $\lambda$ has been introduced to facilitate the discussion, the Hamiltonian can be expressed as
\begin{equation}\label{Eq:16}
\hat{H}'_{S}=\hat{H}_{0}+\lambda\hat{H}_{I},
\end{equation}
where $\hat{H}_{0}=\Delta_{q}\hat{J}_{z}+\widetilde{\omega}_{m}\hat{m}^{\dagger}\hat{m}$ and $\lambda\hat{H}_{I}=G(\hat{J}_{+}+\hat{J}_{-})(\hat{m}+\hat{m}^{\dagger})$. Absolutely, the effective Hamiltonian can also be derived by expanding in terms of $\lambda$ as $\hat{H}_{\rm eff}^{(n)}=\sum_{n'=0}^{n}\lambda^{n'}\hat{H}_{\rm eff}^{[n']}$. We move the Hamiltonian to the interaction framework and choose the appropriate generator of SW transformation to ensure that the first-order Hamiltonian is zero, which enables the derivation of the second-order effective Hamiltonian in the dispersion regime as follows
\begin{equation}\label{Eq:17}
\begin{aligned}
\hat{H}_{\rm eff}^{(2)}(t)\approx&\frac{i}{2}[\hat{S},\hat{H}_{SI}]\\
		\approx&\frac{2G^{2}}{\Delta_{+}}\hat{J}_{z}+\frac{2G^{2}(\Delta_{+}+\Delta_{-})}{\Delta_{+}\Delta_{-}}\hat{J}_{z}\hat{m}^{\dagger}\hat{m}+\frac{G^{2}(\Delta_{+}-\Delta_{-})}{\Delta_{+}\Delta_{-}}\hat{J}_{+}\hat{J}_{-}\\
		&+\frac{G^{2}}{\Delta_{+}+\Delta_{-}}e^{-i(\Delta_{+}-\Delta_{-})t}\hat{J}_{z}\hat{m}^{2}+\frac{G^{2}}{\Delta_{+}+\Delta_{-}}e^{i(\Delta_{+}-\Delta_{-})t}\hat{J}_{z}\hat{m}^{\dagger2}\\
  &-\frac{G^{2}}{2(\Delta_{+}-\Delta_{-})}e^{i(\Delta_{+}+\Delta_{-})t}\hat{J}_{+}^{2}-\frac{G^{2}}{2(\Delta_{+}-\Delta_{-})}e^{-i(\Delta_{+}+\Delta_{-})t}\hat{J}_{-}^{2},\\
\end{aligned}
\end{equation}
where $\hat{S}$ serves as the generator of SW transformation that necessitates iterative solving to ensure that the effective Hamiltonian exhibits block-diagonal in relation to the Hilbert space of the target spin. Obviously, the spin-spin interaction stemming from the magnon arises exclusively within the dispersion regime. Once we consider the magnetic oscillator in the vacuum state and choose specific parameters to satisfy the condition $G^{2}\ll 2(\Delta_{+}+\Delta_{-})(\Delta_{+}-\Delta_{-})=4\Delta_{q}\widetilde{\omega}_{m}$, the time-independent effective Hamiltonian can be written as (more detailed derivations are reported in Appendix \ref{APPENDIX:B})
\begin{equation}\label{Eq:18}
    \hat{H}_{\rm eff}^{(2)}\approx\frac{2G^{2}\Delta_{q}}{\Delta_{+}\Delta_{-}}\hat{J}_{z}+\chi\left( -\hat{J}_{z}^{2}+\mathbf{\hat{J}}^{2}\right) ,
\end{equation}
with $\chi=6G^{2}\widetilde{\omega}_{m}/(\Delta_{+}\Delta_{-})$. Furthermore, as established in Sec. \ref{Sec:II}, the squeezing parameter $r$ can be effectively increased by adjusting the drive field frequency under steady-state conditions, resulting in an exponential increase in the effective spin-spin interaction strength $\chi$.

\subsection{Preparation of the GHZ states}
The preparation of multi-particle GHZ states with high fidelity is of great significance for quantum information processing \cite{62PhysRevLett.65.1838,63PhysRevA.54.R4649}. In terms of the OAT interaction given in Eq. (\ref{Eq:18}), it is possible to achieve GHZ states within one step. Considering the representation of $\hat{J}_{x}$, it is recognized that the multi-qubit excited state (ground state) of the spin is denoted as $\ket{\pm\pm...\pm}$, where $\ket{\pm}=(\ket{e}\pm\ket{g})/\sqrt{2}$. On the other hand, the multi-qubit state $\ket{\pm\pm...\pm}$ can be represented by the collective state $\ket{N/2, \pm N/2}_{x}$. Furthermore, the collective state can also be expressed in terms of the eigenstates of the $\{\mathbf{\hat{J}}^{2}, \hat{J}_{z}\}$ as follows
\begin{subequations}\label{Eq:19}
\begin{align}
&\ket{\frac{N}{2},-\frac{N}{2}}_{x}=\sum_{k=-N/2}^{N/2}C_{k}\ket{\frac{N}{2},k}_{z},\\
&\ket{\frac{N}{2},\frac{N}{2}}_{x}=\sum_{k=-N/2}^{N/2}C_{k}(-1)^{N/2-k}\ket{\frac{N}{2},k}_{z},
\end{align}   
\end{subequations}
with $N$ representing the total spin number and $k$ denoting the spin number in the excited state. Indeed, the eigenstates $\ket{N/2,k}_{z}$ satisfy the following relations: $\mathbf{\hat{J}}^{2}\ket{N/2, k}_{z} = N/2(N/2+1)\ket{N/2, k}_{z}$ and $\hat{J}_{z}\ket{N/2, k}_{z}=k\ket{N/2, k}_{z}$. Considering such special case, we can investigate the preparation of entangled states. Furthermore, assuming the initial state as $\ket{N/2, -N/2}_{x}=\ket{--...-}$, at the evolution time $T=\pi/2\chi$ \cite{58Li2019}, we can obtain the GHZ states 
\begin{equation}\label{Eq:20}
    \begin{aligned}
        \ket{\psi(T)}=\ &\mathcal{U}(T)\ket{\frac{N}{2},-\frac{N}{2}}_{x}\\
        =\ &\mathcal{U}(T)\sum_{k=-N/2}^{N/2}C_{k}\ket{\frac{N}{2},k}_{z}\\
        =\ &\mathcal{A}\sum_{k=-N/2}^{N/2}C_{k}\exp \left({i\frac{\pi}{2}k^{2}}\right) \ket{\frac{N}{2},k}_{z},\\
    \end{aligned}
\end{equation}
where
\begin{equation}\label{Eq:21}
	\mathcal{U}(T)=\exp \left({-i\frac{\Delta_{q}}{\widetilde{\omega}_{m}}\frac{\pi}{2}\hat{J}_{z}}\right) \exp \left({-i\frac{\pi}{2}\mathbf{\hat{J}}^{2}}\right) \exp \left({i\frac{\pi}{2}\hat{J}^{2}_{z}}\right) ,
\end{equation}	
and
\begin{equation}\label{Eq:22}
\mathcal{A}=\exp \left[-i\pi (2\Delta_{q}\hat{J}_{z}/\widetilde{\omega}_{m}+N^{2}/2+N)/4\right] . 
\end{equation}
\begin{figure}[t]
\centering
\subfigure{
\includegraphics[width=0.9\columnwidth]{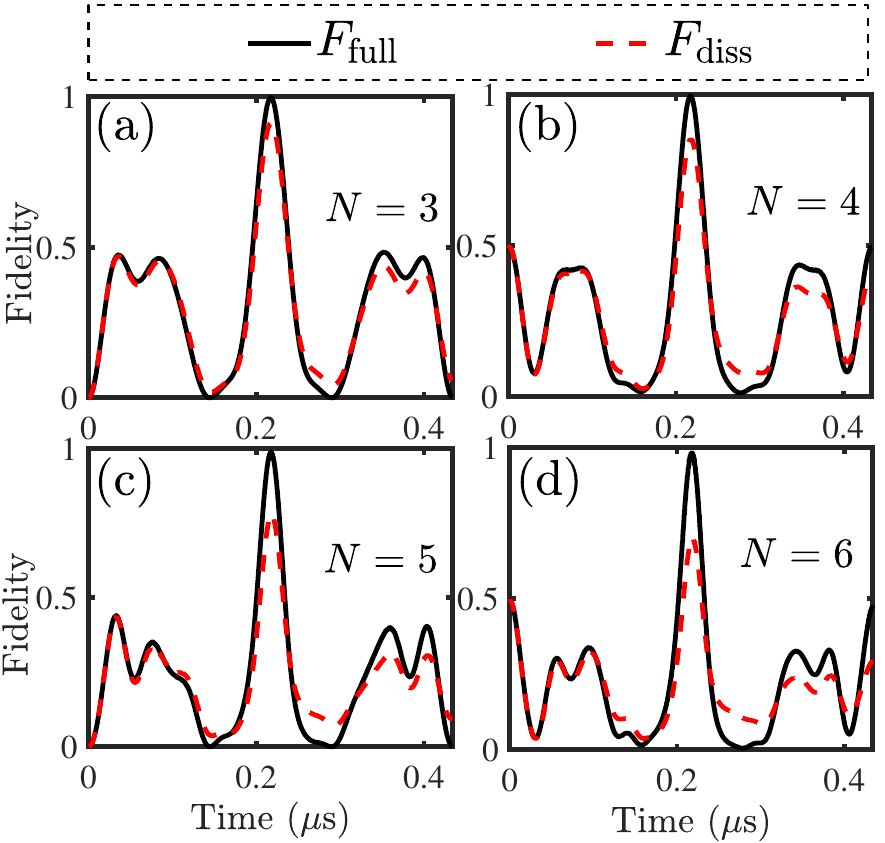}}
\caption{Panels (a)-(d) show the fidelity of evolving state as the a function of the time $t$ in the case of varying spins. The dynamics are governed by the Hamiltonian (\ref{Eq:8}), considering the presence or absence of dissipation. Other parameters are chosen as follows: $\widetilde{\omega}_{m}\approx10G, \Delta_{q}\approx60G, \Delta_{+}\approx70G, \Delta_{-}\approx50G, \gamma_{m}\approx0.005G, \gamma_{q}\approx1.5\times10^{-4}G$, and $r=3$.}
\label{Fig.3}
\end{figure}
Through the aforementioned operation, we have successfully achieved the GHZ states. Next, we delve into the effects of OAT interaction on systems thereby generating GHZ states.

\textit{(i) Even-numbered spins case.} We first begin by examining the scenario with $N$ being even and $k$ as an integer from $-N/2$ to $N/2$. While the outcome of OAT interaction $\exp(i\pi k^{2}/2)$ depends on $k$, specifically given by $e^{i\pi/4}[1+(-1)^{N/2-k}e^{i\pi(N-1)/2}]/\sqrt{2}$. Correspondingly, the GHZ state can be written as
\begin{equation}\label{Eq:23}
    \begin{aligned}
        \ket{\psi(T)}_{e}&=\frac{\sqrt{2}}{2}e^{i\frac{\pi}{4}} \mathcal{A}\left[\ket{--...-}_{x}+i^{N-1}\ket{++...+}_{x}\right].
    \end{aligned}
\end{equation}

\textit{(ii) Odd-numbered spins case.}
In the scenario where $N$ is odd and $k$ ranges from  $-N/2$ to $N/2$ as a half-integer, it may be necessary to introduce an integer $k'=k-1/2$ to ensure $e^{i\pi k^{2}/2}=e^{-i\pi(4k'^{2}+4k-1)/8}$. The GHZ state can be written as 
\begin{equation}\label{Eq:24}
    \begin{aligned}
        \ket{\psi(T)}_{o}=\frac{\sqrt{2}}{2}e^{i\frac{\pi}{4}} \mathcal{A}\left[\ket{--...-}_{x}-i^{N}\ket{++...+}_{x}\right].\\
    \end{aligned}
\end{equation}
It is worth mentioning that applying an additional localized operation $e^{i\pi/8}e^{-i\pi\hat{J}_{z}/2}$ to the above result can formally yield general outcomes of GHZ states without other effects on the ground and excited states in any other way.  According to the above Eq. (\ref{Eq:23}) and Eq. (\ref{Eq:24}), the GHZ states corresponding to an even and odd number of spins at time $t=\pi/2\chi$ have been obtained, respectively. To facilitate a more comprehensive discussion of the proposed scheme, we introduce the dissipation arising from the magnon and spins. With Born-Markov approximation, the time evolution of the density operator $\hat{\rho}$ can be utilized to describe the dynamics of the system as \cite{65Lindblad1976} 
\begin{equation}\label{Eq:25}
\frac{d}{d t} \hat{\rho}= -i[\hat{H}_{S}, \hat{\rho}]+\gamma_{m}\mathcal{L}[\hat{m}] \hat{\rho}+\gamma_{q}\sum^{N}_{j=1}\mathcal{L}[\hat{\sigma}^{-}_{j}] \hat{\rho},
\end{equation}
\begin{figure}[t]
\centering
\includegraphics[width=0.9\columnwidth]{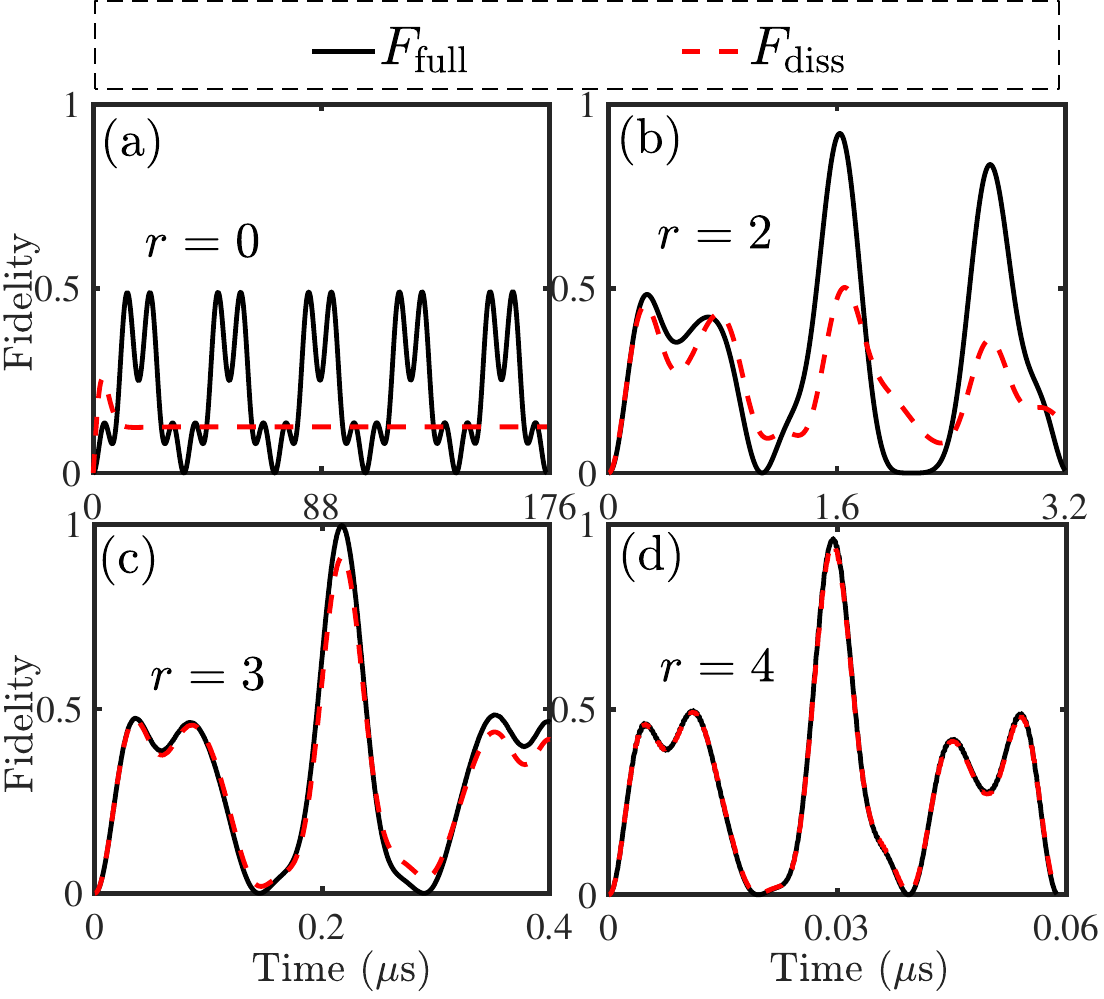}
\caption{Panels (a)-(d) show the fidelity of evolution state over time across various squeezing parameters for $N=3$. The dynamics is governed by the Hamiltonian (\ref{Eq:8}) with regarding to the presence or absence of dissipation. The other parameters are consistent, as in Fig. \ref{Fig.3}.}
\label{Fig.4}
\end{figure}
where the Lindblad operator is $\mathcal{L}[\hat{\mathcal{O}}]\hat{\rho}=\frac{1}{2}(2\hat{\mathcal{O}}\hat{\rho}\hat{\mathcal{O}^{\dag}}-\hat{\rho}\hat{\mathcal{O}^{\dag}}\hat{\mathcal{O}}-\hat{\mathcal{O}^{\dag}}\hat{\mathcal{O}}\hat{\rho})$, $\gamma_{m}$ is the decay rate of magnon mode, $\gamma_{q}$ is relaxation rate of the two level system. In order to evaluate the robustness of the proposal, we need to introduce the definition of fidelity under various spin numbers as: $F\left(\hat{\rho}_{o(e)}, \ket{\psi(T)}_{o(e)}\right)=\bra{\psi(T)}_{o(e)}\hat{\rho}_{o(e)}\ket{\psi(T)}_{o(e)}$ where $\rho_{o(e)}$ is the density matrix at any moment obtained from Eq. (\ref{Eq:25}) for the cases of odd and even spin numbers. Under the dissipative conditions depicted in Fig \ref{Fig.3}, we have detailed the time-dependent dynamics of the GHZ states as they vary with different spin numbers. Obviously, the presence of decoherence has a minimal impact on the fidelity of the target state, indicating that our scheme can effectively mitigate the noise from the external environment to ensure high-fidelity GHZ states. Due to the previous accurate operations, the GHZ states described by Eqs. (\ref{Eq:23}) and (\ref{Eq:24}) exist only at the time $t=\pi/2\chi$, while ones vanish elsewhere. Similarly, the high-fidelity GHZ states can be obtained at time $\chi t=(2Z+1)\pi/2$ with $Z$ being a positive integer. Nevertheless, in comparison to previous findings, the detrimental effects of dissipation gradually accumulate as the dynamical evolution progresses, disrupting the effective coupling between spins and ultimately posing challenges in generating the quantum state. Consequently, summarizing the results from the analysis above, our scheme presents exciting possibilities for high-fidelity GHZ states.

Referring to the aforementioned discussion, it is noted that compression parameters play a pivotal role in amplifying the effective coupling within the system, thereby ensuring a high level of robustness for the prepared GHZ states. Subsequently, we will delve into a detailed exploration of the profound impact exerted by these compression parameters on the dynamic behavior and stability of the system. In Fig. \ref{Fig.4}, we illustrate the fidelity of the target states to verify the robustness of the scheme under varying parameters. Without magnon squeezing (i.e., $r=0$), the term $\hat{H}_{Sq}$ exerts a significant influence on the dynamics of the system, making it difficult to obtain the GHZ states even in a decoherence scenario. Moreover, the effective interaction among the spins is notably weak, and the presence of the dissipation terms nearly completely eliminate the effective coupling between them.
\begin{table}[t]
\centering
\caption{The fidelities of GHZ states with different squeezing parameters $r$ and various values of $\Delta\omega$. The other parameters are consistent as in Fig. \ref{Fig.3}.}
\setlength{\tabcolsep}{5mm}
\begin{tabular}{cccc}
  \hline \hline Fidelity & $r=3$ & $r=3.5$ & $r=4$ \\
\hline $\Delta\omega= 0.00~{\rm MHz}$ & 0.9973 & 0.9864 & 0.9609 \\
 $\Delta\omega =0.18~{\rm MHz}$ & 0.9564 & 0.9799 & 0.9600 \\
 $\Delta\omega =0.36~{\rm MHz}$ & 0.8482 & 0.9581 & 0.9582 \\
 $\Delta\omega =0.54~{\rm MHz}$ & 0.5983 & 0.9275 & 0.9513 \\
 $\Delta\omega =0.73~{\rm MHz}$ & 0.4663 & 0.9236 & 0.9448 \\
 $\Delta\omega =1.09~{\rm MHz}$ & 0.3001 & 0.8168 & 0.9344 \\
 $\Delta\omega =1.82~{\rm MHz}$ & 0.0927 & 0.4639 & 0.8996 \\
\hline \hline\label{Tab1}
\end{tabular}
\end{table}
Additionally, upon comparing the results of each curve, it is evident that variations in the squeezing parameter can substantially impact the fidelity of the target states. In essence, our scheme exhibits excellent robustness to this disturbance. Fortunately, we further shorten the time required to derive the GHZ states, effectively avoiding additional operations and interference within the system.

\section{Eliminate the Effect of Inhomogeneous Broadening}
\label{Sec:IV}
\begin{figure}[t]
\centering
\includegraphics[width=1\columnwidth]{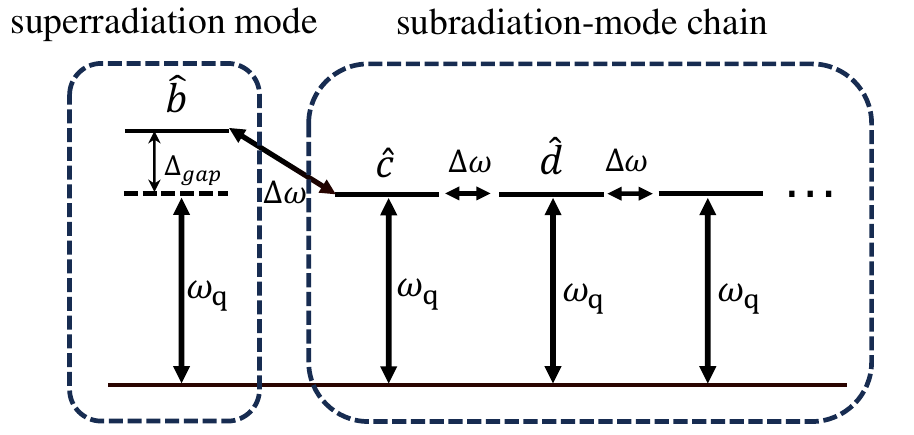}
\caption{Schematic diagram of the coupling between superradiative and subradiative modes. $\Delta_{\rm{gap}}$ is the energy shift generated in the superradiant mode due to the interaction between each spin with transition frequency of $\omega_{q}$ and YIG under adiabatic elimination. $\Delta\omega$ is the coupling strength between different modes caused by inhomogeneous broadening.}
\label{Fig.5}
\end{figure}

In the preceding section, our analysis assumes a consistent frequency for each spin excitation, however, practical scenarios introduce various factors leading to discrepancies in the transition frequencies of individual spins. Consequently, this results in varying degrees of broadening within the absorption spectrum, known as inhomogeneous broadening \cite{59PhysRevA.83.053852}. This disparity in broadening serves as a primary source of typical noise in spin ensembles, ultimately compromising the fidelity of entangled states under preparation. To mitigate the impact of inhomogeneous broadening, strategies such as employing spin echo pulse sequences or adopting the cavity protection method proposed in this section can be implemented effectively. In the thermodynamic limit scenario, by leveraging the Holstein-Primakoff transformation, the localized operators can be represented as
\begin{equation}\label{Eq:26}
\begin{aligned}
\hat{\sigma}_{j}^{z}&=\hat{a}_{j}^{\dag}\hat{a}_{j}-\frac{1}{2},\\
   \hat{\sigma}_{j}^{+}&=\hat{a}^{\dag}_{j}{(1-\hat{a}_{j}^{\dag}\hat{a}_{j})^{\frac{1}{2}}},\\
   \hat{\sigma}^{-}_{j}&={(1-\hat{a}^{\dag}_{j}\hat{a}_{j})^{\frac{1}{2}}}\hat{a}_{j},
\end{aligned}
\end{equation}
where $\hat{a}_{j}$ $(\hat{a}^{\dagger}_{j})$ denotes the creation (annihilation) operator of $j$-th bosonic mode. For the low-lying bosonic excitations, i.e., $\langle\hat{a}_{j}^{\dag}\hat{a}_{j}\rangle\ll 1$, we can safely approximate the localized operators as the following relations: $\hat{\sigma}^{+}_{j}\approx\hat{a}_{j}^{\dag}$, $\hat{\sigma}^{-}_{j}\approx\hat{a}_{j}$. To ensure consistency with the approach used in previous contexts, we introduce the collective bosonic operator $\hat{b}^{\dagger}=\sum_{j=1}^{N}\hat{a}_{j}^{\dagger}/\sqrt{N}$ as the superradiant mode. Then, the Hamiltonian (\ref{Eq:15}) is given by 
\begin{equation}\label{Eq:27}
\hat{H}'=\widetilde{\omega}_{m}\hat{m}^{\dagger}\hat{m}+\sum_{j=1}^{N}\omega_{q_{j}}\hat{a}_{j}^{\dagger}\hat{a}_{j}+G(\hat{b}^{\dagger}+\hat{b})(\hat{m}+\hat{m}^{\dagger}),
\end{equation}
where $\omega_{q_{j}}$ denotes the energy level
splitting of each bosonic mode. To clearly illustrate the effect of inhomogeneous broadening, we have regarded several spins as an example to show the fidelity of the target state under the condition of inhomogeneous detuning, where such inhomogeneity denotes $\Delta\omega=\sum_{j=1}^{N}\delta_{j}/\sqrt{N}$ with $\delta_{j}=|\omega_{q_{j}}-\omega_{q}|$. In Table \ref{Tab1}, we have readily observed that a rise in non-uniform deviations scarcely triggers reliable results for GHZ states. In general, a spectrally narrow ensemble exhibits strong coupling to a magnon, where excitations undergo coherent Rabi oscillations between the superradiant spin-wave mode and the magnon without involving the subradiant modes. In fact, the inhomogeneity in transition frequencies may gradually introduce mixing with the subradiant modes, ultimately resulting in decoherence of the magnon-superradiant subspace. In other words, the spin dephasing processes induced by the inhomogeneous broadening have to bridge the energy gap between the superradiant modes and the subradiant modes. Consequently, we investigate how the energy gap induced by the magnon $\Delta_{\rm{gap}}$ prevents the mixing of the superradiant and subradiant modes. Then, the Hamiltonian following adiabatic elimination can be written as
\begin{equation}\label{Eq:28}
\hat{H}''=(\widetilde{\omega}_{m}-\Delta)\hat{m}^{\dagger}\hat{m}+\sum_{j=1}^{N}\omega_{q}\hat{a}_{j}^{\dagger}\hat{a}_{j}+\Delta_{\rm{gap}}\hat{b}^{\dagger}\hat{b},
\end{equation}
where $\Delta_{\rm{gap}}=2G^{2}\omega_{q_{j}}/(\omega_{q_{j}}-\widetilde{\omega}_{m})(\omega_{q_{j}}+\widetilde{\omega}_{m})$ and $\Delta=2G^{2}\widetilde{\omega}_{m}/(\omega_{q_{j}}-\widetilde{\omega}_{m})(\omega_{q_{j}}+\widetilde{\omega}_{m})$. Furthermore, we can obtain the Heisenberg equation of motion \cite{66Dubbers2013} for $\hat{b}^{\dagger}$ and $\hat{c}^{\dagger}$ as
\begin{equation}\label{Eq:29}
        \dot{\hat{b}}        =i(\omega_{q}+\Delta_{\rm{gap}})\hat{b}^{\dagger}+i\Delta\omega\hat{c}^{\dagger},
\end{equation}
\begin{equation}\label{Eq:30}
     \dot{\hat{c}} =i(\omega_{q}+\Delta_{\rm{gap}})\hat{c}^{\dagger}+i\Delta\omega\hat{d}^{\dagger},
\end{equation}
where $\hat{c}=(\Delta\omega)^{-1}\sum_{j=1}^{N}\delta_{j}\hat{a}_{j}$ and $\hat{d}=(\Delta\omega)^{-2}\sum_{j=1}^{N}\delta_{j}^{2}\hat{a}_{j}$ represent the subradiation mode and the next-to-subradiation mode, respectively (see Fig. \ref{Fig.5}). According to the Heisenberg equation for each radiation mode mentioned above, we can see that such inhomogeneity in the transition frequencies results in interactions between all radiation modes with a coupling intensity of $\Delta\omega$. Due to the interaction between the radiation modes, we must ensure to decouple the radiation modes. For a large gap ($\Delta_{\rm{gap}}\gg \Delta\omega$), the inhomogeneous broadening can not induce real transitions involving the superradiantmode. This energy gap may efficiently protect the quantum information stored in the superradiant spin-wave mode from the dephasing effects of inhomogeneous broadening or spin diffusion. Only in this way, can we evade various mode mixing and guarantee the realization of high-fidelity GHZ states.

\section{Conclusion and Discussion}
\label{Sec:V}
To assess the feasibility of GHZ states generation scheme in a realistic experiment, we now discuss the relevant achievable parameters. For the feasible experimental conditions, we consider the energy level splitting of spins $\Delta_{q}=2\pi\times2.35$ GHz \cite{Wang2023},  the angular frequency of magnon roughly equals as $\widetilde{\omega}_{m}=2\pi\times0.35$ GHz \cite{PhysRevA.109.022442}, the coupling strength between the single magnetic oscillator and the spin is $g=2\pi\times0.69$ MHz \cite{Wang2023}. When considering the dissipation, the decay rates $\gamma_{m}=2\pi\times1$ MHz \cite{PhysRevA.108.063703}, and $\gamma_{q}=2\pi\times1$ kHz \cite{42PhysRevB.105.245310}. Thus, all the necessary conditions for achieving effective dynamics have been successfully satisfied, which  allows for the production of extremely reliable GHZ states at a distance.

In summary, we have proposed a theoretical method for generating high-fidelity GHZ states. This method has primarily utilized the Kerr effect of magnon to enhance the effective coupling strength, which has been a key step in this paper. Currently, experiments have demonstrated the potential existence of spin-orbit coupling in YIG, which can lead to magnetocrystalline anisotropy and induce the Kerr effect \cite{67PhysRevB.94.224410}. This result has provided an experimental basis for our method. Based on this, we have proposed a hybrid system that features direct weak coupling between a magnon and multiple NV centers. Under the action of a strong driving field \cite{55PhysRevB.108.024105,56PhysRevB.107.144411,61PhysRevLett.98.030405}, the Kerr nonlinear term of the magnon has been able to be linearized, ensuring efficient magnon squeezing and resulting in exponentially enhanced magnon-spin coupling. Specifically, we have discussed the relationship between the squeezing parameter $r$ and the driving field \cite{PhysRevA.108.023503}. In order to obtain spin-spin interactions, we have chosen to induce remote spin-spin interactions through adiabatic elimination of the magnon in the dispersive regime. Then we have used the obtained OAT term to prepare multi-particle GHZ states at the specific time \cite{33PhysRevLett.87.230404,58Li2019}. Due to the enhancement of the effective coupling strength, the prepared GHZ states have been able to maintain high fidelity even in the case of dissipation in the system. Finally, we have analyzed the coupling between radiation modes caused by different transition frequencies of the spins, which could reduce the fidelity of the entangled states we have prepared. To address this, we have introduced a cavity protection mechanism to mitigate inhomogeneous broadening effects. Specifically, we have achieved decoupling between the radiation modes using the energy level shifts induced by the magnon \cite{59PhysRevA.83.053852}, ensuring that the GHZ states we have prepared are not significantly affected even with different transition frequencies. Our method has offered a potential possibility for realizing long-distance quantum information processing between spins.

\section*{Acknowledgments}
The author would like to sincerely thank Wuji Zhang for the writing and revision of this article. The work is supported by the scientific research Project of the Education Department of Jilin Province (Grant Nos. JJKH20241408KJ and JJKH20231293KJ). C.W. is supported by the National Research Foundation, Singapore and A*STAR under its Quantum Engineering Programme (NRF2021-QEP2-02-P03).

\appendix
\renewcommand{\appendixname}{APPENDIX}
\section{Linearization of Kerr term}\label{APPENDIX:A}
A powerful tool for describing the time dynamics of open
nonlocal systems is the quantum master equation or its equivalent, the Heisenberg-Langevin equation. We adopt the latter for our current paper, which is the most commonly used method to handle continuous variable systems. According to the Heisenberg-Langevin method, the dynamics of the system can be described by the following nonlinear equations, i.e.
\begin{equation}\label{Eq:A1}
    \begin{aligned}
               &\dot{\hat{m}}=-i(\delta_{m}-i\frac{\gamma_{m}}{2})\hat{m}+iK\hat{m}^{\dagger}\hat{m}^{2}-ig\hat{J}_{-}-i\Omega_{d}+\hat{M}_{\rm{in}},
    \end{aligned}
\end{equation}
where $\hat{M}_{\rm{in}}$ represents the input noise operators induced by the environments. These noise operators have zero mean values, namely, $\braket{\hat{M}_{\rm{in}}}=0.$ 
Next we derive the linearized dynamics of the quantum fluctuations around the steady-state expectation values
of the coupled system. This requires the system to be
driven by a strongly pumped laser, so that the operators in
Eq. (\ref{Eq:A1}) can be decomposed as the sum of the steady-state value and a small fluctuation, i.e., $\hat{A}=\braket{\hat{A}}+\delta\hat{A}$. 
Eq. (\ref{Eq:A1}) can be separated into two parts, 
the average value part
\begin{equation}\label{Eq:A2}
    \begin{aligned}
        &\braket{\dot{\hat{m}}}=-i(\delta_{m}-i\frac{\gamma_{m}}{2})\braket{\hat{m}}+iK|\braket{\hat{m}}|^{2}\braket{\hat{m}}-ig\braket{\hat{J}_{-}}-i\Omega_{d}.
    \end{aligned}
\end{equation}
and the quantum fluctuations part
\begin{equation}\label{Eq:A3}
    \begin{aligned}
        \delta\dot{\hat{m}}=&-i(\delta_{m}-i\frac{\gamma_{m}}{2})\delta\hat{m}-ig\delta\hat{J}_{-}+\hat{M}_{\rm{in}}\\
        &+iK\left( \delta\hat{m}^{\dagger}\delta\hat{m}\delta\hat{m}+2\braket{\hat{m}}\delta\hat{m}^{\dagger}\delta\hat{m}+\braket{\hat{m}}^{2}\delta\hat{m}^{\dagger}\right. \\
        &\left. +\braket{\hat{m}^{\dagger}}\delta\hat{m}\delta\hat{m}+2|\braket{\hat{m}}|^{2}\delta\hat{m}\right) ,
    \end{aligned}
\end{equation}
For the Eq. (\ref{Eq:A2}), we focus on the steady-state solution, which can be obtained as the system reaches a steady state
\begin{equation}\label{Eq:A4}
    \begin{aligned}
        &0=-i(\delta_{m}-i\frac{\gamma_{m}}{2})\braket{\hat{m}}+iK|\braket{\hat{m}}|^{2}\braket{\hat{m}}-ig\braket{\hat{J}_{-}}-i\Omega_{d}.
    \end{aligned}
\end{equation}
In terms of Eq. (\ref{Eq:A4}), the steady-state solutions of the dynamical variables are as follows
\begin{equation}\label{Eq:A5}
    \begin{aligned}
        \braket{\hat{m}}=\frac{\Omega_{d}+g\braket{\hat{J}_{-}}}{K|\braket{\hat{m}}|^{2}-\delta_{m}+i\frac{\gamma_{m}}{2}}.
    \end{aligned}
\end{equation}
Furthermore, it is noteworthy that under strong driving conditions, $ g\braket{\hat{J}_{-}}\ll \Omega_{d}$ can be satisfied, allowing the steady-state average value of $\braket{\hat{m}}$ to be simply expressed as
\begin{equation}\label{Eq:A6}
    \braket{\hat{m}}\approx\frac{\Omega_{d}}{K|\braket{\hat{m}}|^{2}-\delta_{m}+i\frac{\gamma_{m}}{2}}.
\end{equation}
It is observed that the steady-state average value of the system is particularly large, which results in an extremely weak fluctuation amplitude associated with the steady-state value of the operator, which allows to safely disregard higher-order fluctuation components, so we can obtain the quantum Langevin equations for the fluctuation operator
\begin{equation}\label{Eq:A7}
    \begin{aligned}
        \delta\dot{\hat{m}}=&-i(\delta_{m}-i\frac{\gamma_{m}}{2})\delta\hat{m}+2iK|\braket{\hat{m}}|^{2}\delta\hat{m}+iK\braket{\hat{m}}^{2}\delta\hat{m}^{\dagger}\\
        &-ig\delta\hat{J}_{-}+\hat{M}_{\rm{in}}.
    \end{aligned}
\end{equation}
In fact, we focus more on the operator component related to the dynamical evolution, it follows
from Eq. (\ref{Eq:A7}) the linearized Hamiltonian under strong field driving can be written as
\begin{equation}\label{Eq:A8}	\hat{H}_{L}=\Delta_{q}\hat{J}_{z}+\Delta_{m}\hat{m}^{\dagger}\hat{m}-\frac{1}{2}(\mathcal{K}\hat{m}^{2}+\mathcal{K}^{*}\hat{m}^{\dagger2})+g(\hat{J}_{+}\hat{m}+\hat{J}_{-}\hat{m}^{\dagger}),
\end{equation}
where $\Delta_{m}=\omega_{m}-\omega_{d}-2K|\braket{\hat{m}}|^{2}$ and $\mathcal{K}=K\braket{\hat{m}}^{2}$. 
Next, we present the relationship between $\mathcal{K}$ and the Rabi frequency associated with the detuning $\delta_{m}$ of the magnon, the driving field $\Omega_{d}$. 
 According to Eq. (\ref{Eq:A6}), we have a cubic nonlinear equation for the shifted frequency $\mathcal{K}=K\braket{\hat{m}}^{2}$
\begin{equation} \label{Eq:A9} \mathcal{K}^{3}+C_{2}\mathcal{K}^{2}+C_{1}\mathcal{K}+C_{0}=0,
\end{equation}
where $C_{0}=-K\Omega_{d}^{2}$, $C_{1}=\delta_{m}^{2}+\gamma_{m}^{2}/4$, and $C_{2}=-2\delta_{m}$.

\section{Detailed Derivation of Eq. (\ref{Eq:18})}\label{APPENDIX:B}
In this appendix, we give a detailed derivation of the second-order effective Hamiltonian $\hat{H}_{\rm eff}^{(2)}$ of Eq. (\ref{Eq:16}) in the main text. A small parameter $\lambda$ is introduced into the Hamiltonian for further discussion, so that Eq. (\ref{Eq:15}) can be written as $\hat{H}'_{S}=\hat{H}_{0}+\lambda\hat{H}_{I}.$ Since we focus on the dynamic evolution process of the system, that is, the impact of the interaction part on the dynamics of the system, we will discuss the next problem in the interaction picture. First take a unitary transformation $\hat{U}(t)=\exp[-i(\Delta_{q}\hat{J}_{z}+\widetilde{\omega}_{m}\hat{m}^{\dagger}\hat{m})t]$ of the Hamiltonian (\ref{Eq:15}) and transform it into the interaction drawing scene with
\begin{equation}\label{Eq:B1}
		\begin{aligned}
			\hat{H}'_{SI}(t) &= \hat{U}^{\dagger}\hat{H}'_{S}\hat{U}-i\hat{U}^{\dagger}\frac{\partial\hat{U}}{\partial t}\\
			&=G\left( \hat{J}_{+}\hat{m}e^{i\Delta_{-}t}+\hat{J}_{+}\hat{m}^{\dagger}e^{i\Delta_{+}t}+\hat{J}_{-}\hat{m}e^{-i\Delta_{+}t}+\hat{J}_{-}\hat{m}^{\dagger}e^{-i\Delta_{-}t}\right) ,\\
		\end{aligned}
	\end{equation}
 we apply a SW transform to get the effective Hamiltonian
 \begin{equation}\label{Eq:B2}
	\hat{H}_{\rm{eff}}(t)=e^{i\hat{S}(t)}[\hat{H}'_{SI}(t)-i\partial t]e^{-i\hat{S}(t)},
\end{equation}
where $\hat{S}(t)$ is the generator in the SW transformation, it can also be represented by the small parameter $\lambda$
\begin{equation} \label{Eq:B3}
\hat{S}(t)=\sum\limits_{n=1}\limits^{\infty}\lambda^{n}\hat{S}_{n}(t).
\end{equation}
Next to obtain the effective Hamiltonian we resort to the BCH formula
\begin{equation}\label{Eq:B4}
    e^{\hat{A}} \hat{B} e^{-\hat{A}}=\sum_{n=0}^{\infty} \frac{1}{n!} \mathcal{C}_{n}[\hat{A}] \hat{B}=\hat{B}+[\hat{A}, \hat{B}]+\frac{1}{2}[\hat{A},[\hat{A}, \hat{B}]]+\cdots,
\end{equation}
where
\begin{equation}\label{Eq:B5}
    \mathcal{C}_{n}[\hat{A}](\bullet)=[\underbrace{\hat{A},[\hat{A},[\hat{A},[\ldots}_{\mathrm{n}}, \bullet]]]] ,
\end{equation}
is a nested commutator. The effective Hamiltonian of Eq. (\ref{Eq:B2}) consists of two parts
    \begin{equation}\label{Eq:B6}
    \begin{aligned}
        e^{i\hat{S}(t)}\hat{H}'_{SI}(t)e^{-i\hat{S}(t)}&=\lambda\hat{H}'_{SI}+i[\hat{S},\lambda\hat{H}'_{SI}]+\frac{i^{2}}{2!}[\hat{S},[\hat{S},\hat{H}'_{SI}]]+\cdots\\
        &=\lambda\hat{H}'_{SI}(t)+\lambda^{2}\left( i[\hat{S}_{1},\hat{H}'_{SI}]\right) +O(\lambda^{3}),\\
    \end{aligned}
    \end{equation}
    and
    \begin{equation}\label{Eq:B7}
    \begin{aligned}
        e^{i\hat{S}(t)}(-i\partial t)e^{-i\hat{S}(t)}&=-\sum\limits_{n=0}^{\infty}\frac{i^{n}}{(n+1)!}C_{n}[\hat{S}]\dot{\hat{S}}\\
        &=-\dot{\hat{S}}-\frac{i}{2!}[\hat{S},\dot{\hat{S}}]+\frac{1}{3!}[\hat{S},[\hat{S},\dot{\hat{S}}]]\cdots\\
        &=-\lambda\dot{\hat{S}}_{1}+\lambda^{2}\left( -\dot{\hat{S}}_{2}-\frac{i}{2}[\hat{S}_{1},\dot{\hat{S}}_{1}]\right) +O(\lambda^{3}).\\
    \end{aligned}
    \end{equation}
    
According to Eq. (\ref{Eq:B6}) and Eq. (\ref{Eq:B7}), we can obtain the effective Hamiltonian of each order of expansion as follows
\begin{subequations}\label{Eq:B8}
\begin{align}
    &\hat{H}_{\rm{eff}}^{(0)}=0, \\
    &\hat{\mathcal{H}}_{\rm{eff}}^{(1)}=-\dot{\hat{S}}_{1}+\hat{H}'_{SI},\\
    &\hat{H}_{ \rm{eff}}^{(2)}=-\dot{\hat{S}}_{2}-\frac{i}{2}\left[\hat{S}_{1}, \dot{\hat{S}}_{1}\right]+i\left[\hat{S}_{1}, \hat{H}'_{SI}\right].
\end{align}
\end{subequations}
In the dispersion region (i.e., $G/\Delta_{\pm}\ll 1$), the effect of the higher-order terms is very small and negligible, so we keep the effective Hamiltonian only to the second order and take condition $\hat{H}_{\rm{eff}}^{(1)}=0$, in which case we can obtain
\begin{equation}\label{Eq:B9}
\begin{aligned}
    		\hat{S}(t)=&\int\hat{H}'_{SI}dt\\
		=&\frac{G}{i\Delta_{-}}\hat{J}_{+}\hat{m}e^{i\Delta_{-}t}+\frac{G}{i\Delta_{+}}\hat{J}_{+}\hat{m}^{\dagger}e^{i\Delta_{+}t}-\frac{G}{i\Delta_{+}}\hat{J}_{-}\hat{m}e^{-i\Delta_{+}t}\\
		&-\frac{G}{i\Delta_{-}}\hat{J}_{-}\hat{m}^{\dagger}e^{-i\Delta_{-}t}.
\end{aligned}
\end{equation}
The second-order effective Hamiltonian is obtained
\begin{equation}\label{Eq:B10}
		\begin{aligned}
		\hat{H}_{\rm eff}^{(2)}(t)\approx&\frac{i}{2}[\hat{S},\hat{H}'_{SI}]\\
		\approx&\frac{2G^{2}}{\Delta_{+}}\hat{J}_{z}+\frac{2G^{2}(\Delta_{+}+\Delta_{-})}{\Delta_{+}\Delta_{-}}\hat{J}_{z}\hat{m}^{\dagger}\hat{m}+\frac{G^{2}(\Delta_{+}-\Delta_{-})}{\Delta_{+}\Delta_{-}}\hat{J}_{+}\hat{J}_{-}\\
		&+\frac{G^{2}}{\Delta_{+}+\Delta_{-}}e^{-i(\Delta_{+}-\Delta_{-})t}\hat{J}_{z}\hat{m}^{2}+\frac{G^{2}}{\Delta_{+}+\Delta_{-}}e^{i(\Delta_{+}-\Delta_{-})t}\hat{J}_{z}\hat{m}^{\dagger2}\\
		&-\frac{G^{2}}{2(\Delta_{+}-\Delta_{-})}e^{i(\Delta_{+}+\Delta_{-})t}\hat{J}_{+}^{2}-\frac{G^{2}}{2(\Delta_{+}-\Delta_{-})}e^{-i(\Delta_{+}+\Delta_{-})t}\hat{J}_{-}^{2},\\
		\end{aligned}
	\end{equation}
 in the dispersion region, condition $G^{2}\ll 2(\Delta_{+}+\Delta_{-})(\Delta_{+}-\Delta_{-})$ can also be satisfied, so that the high-frequency term can be ignored and the magnetic oscillator is initially in the vacuum state. In these cases, Eq. (\ref{Eq:17}) in the main text can be obtained as follows
 \begin{equation}\label{Eq:B11}
		\hat{H}^{(2)}_{\rm eff}\approx\frac{2G^{2}\Delta_{q}}{\Delta_{+}\Delta_{-}}\hat{J}_{z}+\frac{2G^{2}\widetilde{\omega}_{m}}{\Delta_{+}\Delta_{-}}\left( -\hat{J}_{z}^{2}+\hat{\mathbf{J}}^{2}\right) 	.
	\end{equation}

\bibliography{ref}
\providecommand{\noopsort}[1]{}\providecommand{\singleletter}[1]{#1}%
\end{document}